\documentclass[twocolumn]{aastex631}

\usepackage{amsmath}

\begin{document}

\title{Effect of Finite-Temperature $\beta$-Decay Rates on the Rapid Neutron Capture Process}

\correspondingauthor{Yukiya Saito}
\email{ysaito@nd.edu}

\author[0000-0003-1320-8903]{Yukiya Saito}
\affiliation{Department of Physics and Astronomy, University of Notre Dame, Notre Dame, IN 46556, USA}
\affiliation{Department of Physics and Astronomy, University of Tennessee, Knoxville, Tennessee 37996, USA}
\affiliation{Facility for Rare Isotope Beams, Michigan State University, East Lansing, Michigan 48824, USA}

\author[0000-0001-9639-5382]{Ante Ravli\'{c}}
\affiliation{Facility for Rare Isotope Beams, Michigan State University, East Lansing, Michigan 48824, USA}
\affiliation{Department of Physics, Faculty of Science, University of Zagreb, Bijeni\v cka c. 32, 10000 Zagreb, Croatia}

\author[0009-0000-0107-0244]{Pranav Nalamwar}
\affiliation{Department of Physics and Astronomy, University of Notre Dame, Notre Dame, IN 46556, USA}

\author[0000-0002-4729-8823]{Rebecca Surman}
\affiliation{Department of Physics and Astronomy, University of Notre Dame, Notre Dame, IN 46556, USA}



\begin{abstract}
$\beta$-decay is known to play an essential role in the rapid neutron capture process ($r$-process) during $(n, \gamma) \leftrightarrow (\gamma, n)$ equilibrium and freeze-out when the neutron-rich nuclei decay back to stability. Recent systematic theoretical studies on $\beta$-decay at finite temperature indicated that under hot conditions ($T\sim10$~GK), a significant acceleration of $\beta$-decay rates is expected, especially for nuclei near stability. This corresponds to the early stage of the $r$-process. In this study, we investigate the effect of the $\beta$-decays in finite temperature using the rates calculated with the finite-temperature proton-neutron relativistic quasiparticle random-phase approximation (FT-PNRQRPA).
We explore a variety of astrophysical conditions and find that the effect on the abundance pattern is significant in hot and moderately neutron-rich conditions such as are expected in magnetorotational supernovae. Accelerated $\beta$-decay rates also increase the heating rate in the early phase, resulting in an additional modification of the final abundance pattern.
\end{abstract}



\section{Introduction} \label{sec:intro}
The crucial role of $\beta$-decay in the synthesis of heavy elements has been well recognized since the seminal works by \cite{BBFH} and \cite{Cameron}. The astrophysical slow, intermediate, and rapid neutron capture processes ($s$-, $i$- and $r$-process, respectively), refer to the rate of neutron capture with respect to the typical rate of $\beta$-decay in the regions where each process operates. 

In the $r$-process, the $\beta$-decay rates determine how fast material transfers from one isotopic chain to another during the $(n,\gamma)\leftrightarrow (\gamma, n)$ equilibrium. During this phase, the relative abundances of each isotopic chain are determined by the effective $\beta$-decay rates. {This $\beta$-flow equilibrium approximation provides} 
an explanation for the origin of the second and third $r$-process peaks, reflecting the longer $\beta$-decay half-lives along the $N=82$ and 126 shell closures \citep{Horowitz2019, Cowan2021}. In the late phase of the $r$-process when the material decays back to stability (freeze-out), the neutron capture rates become similar to those of the $\beta$-decays. This competition affects the details of the final abundance pattern, especially in the rare-earth peak (REP) region \citep{Surman_rep_1997,Mumpower_rep_2012a}. $\beta$-decay, along with fission {and alpha decay}, powers the electromagnetic emission from an $r$-process event such as neutron star merger, providing opportunities for detecting the evidence of the $r$-process as it happens (see e.g. \cite{Shibata2019,Metzger2019_kilonovae,Barnes2020_kilonovae} for reviews).

{Recent advances in isotope production capabilities and experimental methods have resulted in impressive gains in the number of neutron-rich species with measured half-lives \citep{Lurosso2015, Wu2017, Wu2020, Kiss2022,Yokoyama+2023,Xu+2023,Tolosa-Delgado+2025}.}
{Still, }many of the neutron-rich nuclei relevant to the $r$-process have not yet been experimentally studied. Therefore, any $r$-process calculation must rely on theoretical predictions of nuclear properties. For $\beta$-decay, there has been remarkable progress in global theoretical descriptions: (semi-)gross theory (\cite{Takahashi1969, Nakata1997}), quasiparticle random-phase approximation (QRPA) with finite-range droplet model (FRDM) \citep{Moller1997,Moller2003,Moller2019}, QRPA with relativistic density functional theory (DFT)\citep{Niu2013, Marketin_RQRPA_2016}, and finite-amplitude method with Skyrme DFT \citep{Ney2020}. The effect of global $\beta$-decay properties on the $r$-process has been a topic of recent investigations \citep{Niu2013, Lund2023, Chen2023}.

The nuclear shell model has previously been used to investigate the effect of stellar environment on astrophysical weak-decay rates (electron capture and $\beta$-decay). \cite{Langanke2000} performed calculations for around 100 nuclei in the $pf$-shell region for the temperature range of 1 -- 10 GK. More recently, \cite{Zha2019} performed calculations employing the improved pf-GXPF1J interaction for a similar mass and temperature range. These studies included only the contribution of allowed transitions, however, in neutron-rich nuclei, first-forbidden transitions tend to play a significant role. Although shell model calculations can provide accurate description of the underlying strength functions in reproducing experimental data, they become computationally unfeasible for global calculations required for studying the $r$-process. A more suitable microscopic approach to such global calculations is DFT, which offers manageable scalability with the system size.

Recently, the relativistic nuclear energy density functional theory has been extended to include the finite-temperature effect in the $\beta$-decay calculation, referred to as the finite-temperature proton-neutron relativistic quasiparticle random-phase approximation (FT-PNRQRPA) \citep{Ravlic2021}. Previously, the finite-temperature effect on the $\beta$-decay half-lives was also studied within the Skyrme-Hartree Fock + Bardeen-Cooper-Schrieffer framework, though only considering the Gamow-Teller excitations \citep{Minato2009}. In their study, it was pointed out that the thermal effect on the $\beta$-decay half-life would be negligible at 0.2~MeV for even-even $N = 82$ isotones. Temperature-dependence of $\beta$-decay was also investigated within the relativistic time-blocking approximation (RTBA), including complex configurations going beyond the QRPA \citep{Litvinova2020a} for doubly-magic ${}^{48}$Ca, ${}^{78}$Ni and ${}^{132}$Sn. In these nuclei, the temperature effects begin to play a role for $T > 6$~GK, consistent with the results reported in \cite{Ravlic2021} and \cite{Minato2009}. Considering that the temperature of the stellar environment in which the $r$-process occurs can reach 10~GK ($\sim 0.9$~MeV) and above, acceleration of $\beta$-decay rates above 6~GK ($\sim 0.5$~MeV) for various nuclei may have a considerable effect on the outcome of nucleosynthesis.  

In this study, taking advantage of the new calculations based on the FT-PNRQRPA framework, {we investigate }the impact of the finite-temperature $\beta^-$-decay rates of the nuclei with proton numbers $20 \leq Z \leq 60$ on $r$-process nucleosynthesis.
The structure of the current paper is as follows. In Section~\ref{sec:method}, we provide a summary of the FT-PNRQRPA method, other relevant nuclear physics inputs of the nuclear reaction network calculations, and the astrophysical conditions under which the nuclear abundances are evolved. In Section~\ref{sec:results}, we discuss the findings from the $r$-process calculations. In Section~\ref{sec:conclusion}, we summarize the current study.

\section{Theoretical Method} \label{sec:method}
\subsection{Finite-temperature \texorpdfstring{$\beta$}{beta}-decay\label{subsec:method_FT_beta}}
The $\beta$-decay rates in this work are calculated within the framework of relativistic DFT at finite-temperature. The initial nuclear state is calculated using the finite-temperature Hartree-BCS (FT-HBCS) theory. Compared to previous work in \cite{Ravlic2021}, which used a rather simple monopole pairing interaction in the isovector channel, we implement a more sophisticated pairing part of the Gogny D1S interaction \citep{Vretenar2005}, with the overall strength $V_{pair}$ determined by reproducing the pairing gaps in semi-magic isotopic chains. This makes the method more suitable for large-scale calculations. In the particle-hole (\textit{i.e.} mean-field) channel, we employ the momentum-dependent D3C* interaction \citep{Marketin_RQRPA_2016}. We assume spherical symmetry in our calculations, a reasonable approximation at higher temperatures where potential energy surfaces soften \citep{Ravlic2024}. Nuclear excitations are calculated based on the FT-PNRQRPA with details described in 
\cite{Ravlic2021}. We include contributions from both allowed ($1^+$) and first-forbidden ($0^-, 1^-$, and $2^-$) transitions, the latter being especially important for neutron-rich nuclei at higher temperatures. Furthermore, the effects of de-excitations and phase-transition from superfluid to a normal state are included self-consistently. All the numerical considerations are kept the same as in \cite{Ravlic2021}. In particular, the FT-HBCS equations are solved in the harmonic oscillator basis with 20(21) shells for fermion(boson) states. The FT-PNRQRPA matrix is diagonalized explicitly in a two-quasiparticle basis, with the cut-off on the sum of proton(neutron) quasiparticle energies, $E_{\pi(\nu)}$, being $E_\pi + E_\nu < 100$ MeV. The calculated $\beta$-decay rates were validated in \cite{Ravlic2021} against the experimental data at zero-temperature, while at finite-temperature a fair agreement was obtained with shell-model calculations \citep{Langanke2001,Mori2016} for selected $pf$-shell nuclei. We note that improved description of the pairing interaction does not significantly alter the rates compared to \cite{Ravlic2021}.

{We evaluate} the $\beta$-decay rates
on a grid spanning temperatures up to 10 GK and a product of stellar density $\rho$ and electron-to-baryon ratio $Y_e$ --- $\rho Y_e$ --- up to $10^{14}$ g/cm$^3$. The calculations include nuclei with proton number $20 \leq Z \leq 60$, from the valley of stability up to the neutron drip line, which is defined by the condition on neutron chemical potential $\lambda_n < 0$. For odd-$A$ and odd-odd nuclei, we assume false even-even vacuum and restrict the particle number to an odd number of nucleons. 

\begin{figure}
    \centering
    \includegraphics[width=\linewidth]{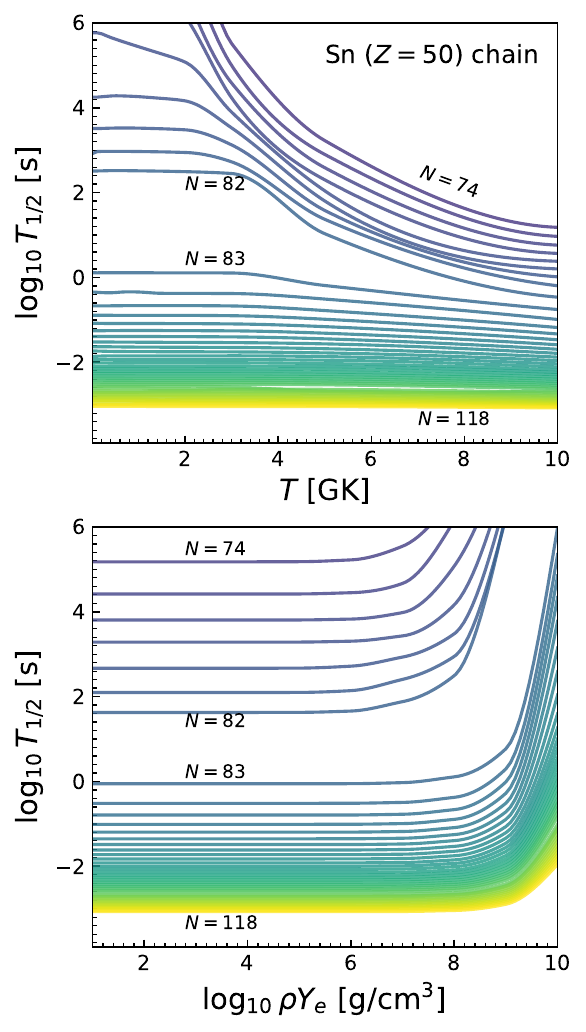}
    \caption{(Upper panel) Temperature evolution of $\beta$-decay half-lives $T_{1/2}$ in tin isotopes ($Z = 50$) for a fixed $\rho Y_e = 10^8$ g/cm${}^3$. (Lower panel) The $\beta$-decay half-lives with respect to density $\rho Y_e$ for a fixed temperature $T = 2$ GK. The neutron numbers of particular tin isotopes are explicitly indicated.}
    \label{fig:temp-dens-dep}
\end{figure}

In the following, we briefly illustrate the main influence of temperature and density-dependence on $\beta$-decay half-lives. This is exemplified in Fig. \ref{fig:temp-dens-dep} for the tin ($Z = 50$) isotopic chain. First, in the upper panel of Fig. \ref{fig:temp-dens-dep}, we fix the density to $\rho Y_e = 10^8$ g/cm${}^3$ and show the temperature evolution of the $\beta$-decay half-lives $T_{1/2}$. A systematic decrease of half-lives with increasing neutron number can be observed, starting from ${}^{124}$Sn to ${}^{168}$Sn. For nuclei near the valley of stability, with longer half-lives at zero temperature, a strong reduction of half-lives with increasing temperature can be noticed, by orders of magnitude. This trend continues until the doubly-magic ${}^{132}$Sn is reached, while for heavier tin isotopes we notice that half-lives are almost temperature independent up to 10 GK. It is interesting to notice the apparent gap in half-lives between ${}^{132}$Sn and ${}^{133}$Sn, due to the shell closure. As the temperature is increased and shell effects weaken, half-lives of ${}^{132}$Sn and ${}^{133}$Sn become more similar.

In the lower panel of Fig. \ref{fig:temp-dens-dep} we fix the temperature to a rather low value of $T = 2$ GK and vary the density $\rho Y_e$. It can be noticed that up to $\rho Y_e = 10^9$ g/cm$^3$, the half-lives are almost independent of $\rho Y_e$, after which they start to increase exponentially. Indeed, as the density increases, the phase-space available for outgoing leptons gets Pauli blocked, decreasing the phase-space integrals and the effective $Q_\beta$-value.  

{In an astrophysical environment with temperature $\sim$10 GK, atoms would be completely ionized. For ionized atoms, bound state $\beta$-decay may become relevant as part of finite-temperature effect \cite{TakahashiYokoi1983}. However, literature discusses the process in the context of the $s$-process nucleosynthesis. In this work, we do not consider the effect of bound state $\beta$-decay; its effect in the context of the $r$-process will be explored in future studies.}

\subsection{Nuclear physics inputs in nuclear reaction network calculation}
In this work, {we compute} the abundance evolution 
using the nuclear reaction network code \textsc{PRISM} \citep{SprouseTheis2020}, which provides a convenient framework for investigating the effect of input from nuclear physics. The nuclear physics inputs in the current calculations are the following. For reactions that involve charged particles ($\alpha$-particle and proton), we adopt the rates from JINA Reaclib \citep{Cyburt2010}. Neutron capture rates are calculated as in \cite{Mumpower2018} using the statistical Hauser-Feshbach (HF) code \textsc{CoH3} with nuclear masses from FRDM2012 \citep{Moller2016} or AME2020 \citep{Wang2021} whenever experimental data exist. Photodissociation rates are calculated within PRISM using detailed balance based on the neutron capture rates and one-neutron separation energies calculated from the same set of theoretical and experimental masses. 
For all types of fission, we adopt symmetric yields without neutron emission to simplify the calculation. Spontaneous fission and alpha decay rates are taken from Nubase2020. Neutron-induced fission rates are calculated with \textsc{CoH3} {as in \cite{Mumpower2018}} using the masses from FRDM2012 or AME2020 and the fission barriers from FRLDM \citep{Moller2015}. $\beta$-delayed fission rates are calculated using the QRPA+HF framework \citep{Mumpower2018}. 

For $\beta^{-}$-decay, which is the focus of the current study, we used the temperature ($T$) and electron density ($\rho Y_e$) dependent $\beta$-decay rates for nuclei with proton numbers $20 \leq Z \leq 60$, calculated with the FT-PNRQRPA framework. The data define a 2D grid of $\beta$-decay rates from $0.01 \leq T~\mathrm{[GK]} \leq 10$ and $10^1 \leq \rho Y_e~\mathrm{[g/cm^3]} \leq 10^{14}$, in total 154 data points, and are linearly interpolated at given $T$ and $\rho Y_e$. Outside the grid, the value at the closest edge of the grid is used. If experimental $\beta$-decay rates are available in Nubase2020, we use them as rates at $T=0$. For the nuclei with temperature-dependent $\beta$-decay rates, we do not consider $\beta$-delayed neutron emission. For all other nuclei, we use the $\beta$-decay rates from \cite{Marketin_RQRPA_2016}. These rates were calculated using the PNRQRPA framework, similar to the current finite-temperature calculations. This is to maintain some degree of consistency in the treatment of $\beta$-decay within the nucleosynthesis calculation.

\subsection{Astrophysical conditions\label{subsec:astro_conditions}}
We consider two types of astrophysical conditions for the $r$-process.
One is a set of ten trajectories (temporal evolution of temperature and baryon density) obtained from the tracer particles in the simulation of a magnetorotational supernova, studied in \cite{Obergaulinger2017, Reichert2021}. In this scenario, the core-collapse supernova of a progenitor with a high rotation rate and a large magnetic field promptly produces neutron-rich polar jet-like ejecta. The prompt ejecta escape interactions with neutrinos, which would render the material less neutron-rich, and elements up to the third $r$-process peak ($A\sim195$) can be synthesized \citep{Cameron2003, Nishimura2006, Winteler2012, Nishimura2015, Nishimura2017, Mosta2018, Reichert2021}. The set of ten trajectories is from \cite{Reichert2021} and the sum of the nucleosynthesis yields from the ten trajectories approximately reproduce the integrated abundance pattern of all tracers in their simulation. \cite{Reichert2021} notes that the integrated abundance pattern does not match the solar abundance pattern well, and they attribute the cause to the uncertainty of nuclear physics inputs, and the possibility of contributions from other sources of $r$-process nucleosynthesis such as neutron star mergers.

The other astrophysical condition is the outflow from the accretion disk formed around a hypermassive neutron star (HMNS) as a result of a neutron star merger. In this scenario, moderately neutron-rich material is ejected from the disk powered by viscous heating and nuclear recombination, which is referred to as a disk wind. Depending on the lifetime of the HMNS produced in a neutron star merger, strong $r$-process elements may be produced \citep{Kaplan2014, Metzger2014, Perego2014, Martin2015, Lippuner2017, Fujibayashi2018}. The set of ten trajectories used in the current study is based on \cite{Metzger2014} with a HMNS lifetime of 100~ms. The ten trajectories were selected using the Actinide-Dilution with Matching (ADM) method, which selects the trajectories to match the calculated abundances of Zr, Dy, and Th with the abundances observed in metal-poor stars \citep{Holmbeck+2023}.

In all astrophysical trajectories employed in this study, the initial composition of the nuclei was calculated using the SFHO equation of state \citep{Steiner2013} at 10~GK based on the density and electron fraction at the moment, assuming that nuclear statistical equilibrium (NSE) is established.

\begin{deluxetable}{lcc}
    \tablecaption{\label{tab:trajectory}Initial conditions of the astrophysical trajectories used in this study. ``MHD'' refers to the magnetorotational supernova from \cite{Reichert2021}, and ``ADM'' refers to the HMNS disk outflow from \cite{Metzger2014} with the trajectories selected with the ADM method \citep{Holmbeck+2023}. The superscript ``init'' corresponds to the condition at $T=10$GK, which is when the nucleosynthesis calculation starts. }
    \tablehead{
        \colhead{Trajectory} & \colhead{$\rho^{\mathrm{init}}\times 10^8$} & \colhead{${Y_e}^{\mathrm{init}}$} \\
        {} & \colhead{[g/cm$^{3}$]} & {}
    }
    \startdata
        MHD\_1 & 1.82 & 0.233\\
        MHD\_2 & 2.27 & 0.271\\
        MHD\_3 & 2.70 & 0.175\\
        MHD\_4 & 3.78 & 0.202\\
        MHD\_5 & 3.23 & 0.238\\
        MHD\_6 & 0.780 & 0.292\\
        MHD\_7 & 2.63 & 0.241\\
        MHD\_8 & 1.56 & 0.230\\
        MHD\_9 & 0.561 & 0.245\\
        MHD\_10 & 1.54 & 0.228\\
        ADM\_1 & 0.454 & 0.299\\
        ADM\_2 & 0.224 & 0.335\\
        ADM\_3 & 0.593 & 0.292\\
        ADM\_4 & 0.468 & 0.287\\
        ADM\_5 & 0.597 & 0.287\\
        ADM\_6 & 0.646 & 0.290\\
        ADM\_7 & 3.13 & 0.123\\
        ADM\_8 & 2.61 & 0.137\\
        ADM\_9 & 0.304 & 0.401\\
        ADM\_10 & 1.80 & 0.150\\
    \enddata
\end{deluxetable}

\section{Results and Discussion} \label{sec:results}
In this section, we report the results of the nucleosynthesis calculations with the two sets of hydrodynamic trajectories described above and in Table I.
Since our finite-temperature $\beta$-decay ($\beta^{\mathrm{FT}}$-decay) rates exist for nuclei with proton numbers $20 \leq Z \leq 60$, we first focus our analysis on the weak (limited) $r$-process, mainly synthesizing nuclei up to the second $r$-process peak at $A \sim 125$. We then discuss the effects on main $r$-process nucleosynthesis.

\subsection{Effect on the weak \textit{r}-process abundances \label{subsec:effect_weak-r}}
\begin{figure*}[ht!]
\includegraphics[width=\textwidth]{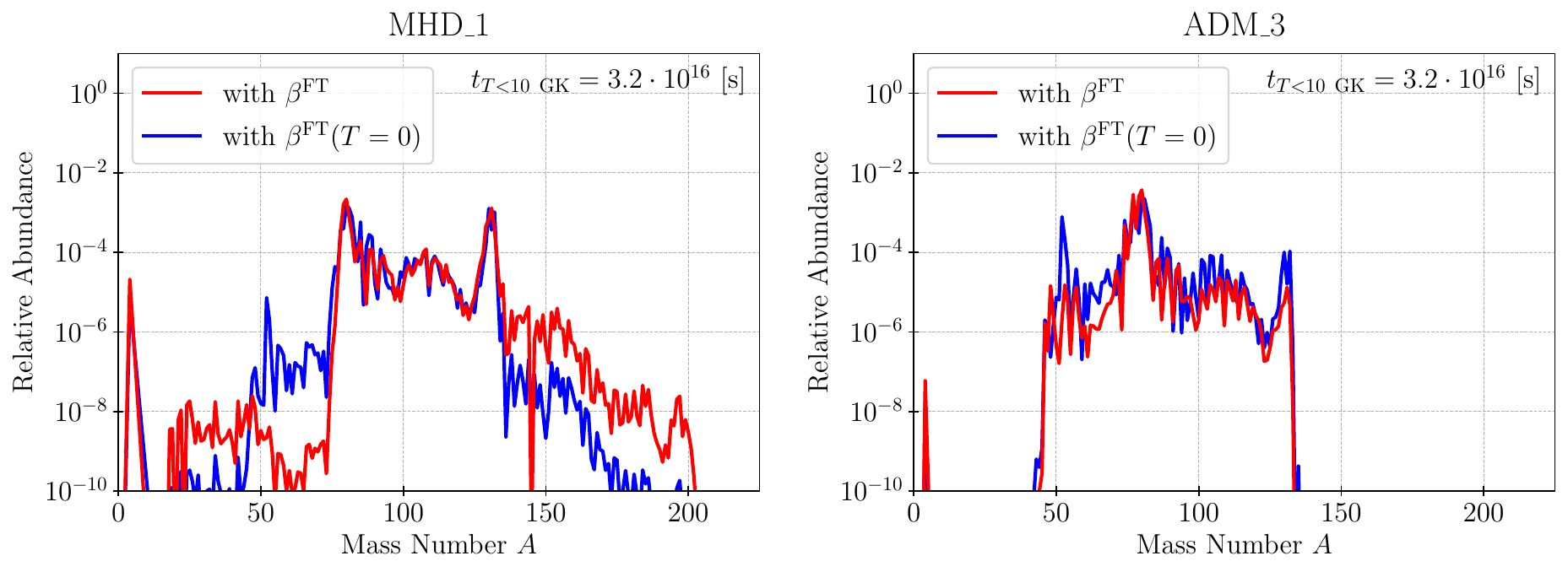}
\caption{
Isotopic abundance patterns (abundances as functions of the mass number $A$) at 1 Gyr to for the MHD\_1 (left) and ADM\_3 trajectories (right). Red color indicates calculations with temperature-dependent $\beta$-decay rates ($\beta^{\mathrm{FT}}$), while blue color represents zero-temperature rates ($\beta^{\mathrm{FT}}(T=0)$).
\label{fig:YA_final}}
\end{figure*}

Figure~\ref{fig:YA_final} shows the comparisons of isotopic abundance patterns at 1~Gyr, computed with the $\beta^{\mathrm{FT}}$-decay rates (red) and with the rates fixed with the values at zero temperature ($\beta^{\mathrm{FT}} (T=0)$, blue) for the MHD\_1 (left) and ADM\_3 trajectories. 
We choose these trajectories as examples of weak $r$-process conditions in the MHD and ADM sets, respectively.
It can be seen in the figure that in the MHD\_1 trajectory, the heavier tail of the abundance pattern is enhanced when the temperature-dependent rates are employed, while in the ADM\_3, the heavy end of the abundance pattern is suppressed. These apparently contradictory effects are due to the complex interplay between the faster temperature-dependent $\beta$-decay rates and neutron capture. The faster high-temperature beta decay rates tend to accelerate the $r$-process flow and ostensibly push material to higher $A$. This, however, results in a more rapid consumption of neutrons, of which there is only a limited supply in a weak $r$-process. The resulting abundance pattern can therefore be either truncated or enhanced in $A$ depending on this interplay, as shown in Figure~\ref{fig:YA_final} and as detailed in the following.

\begin{figure*}[ht!]
\includegraphics[width=\textwidth]{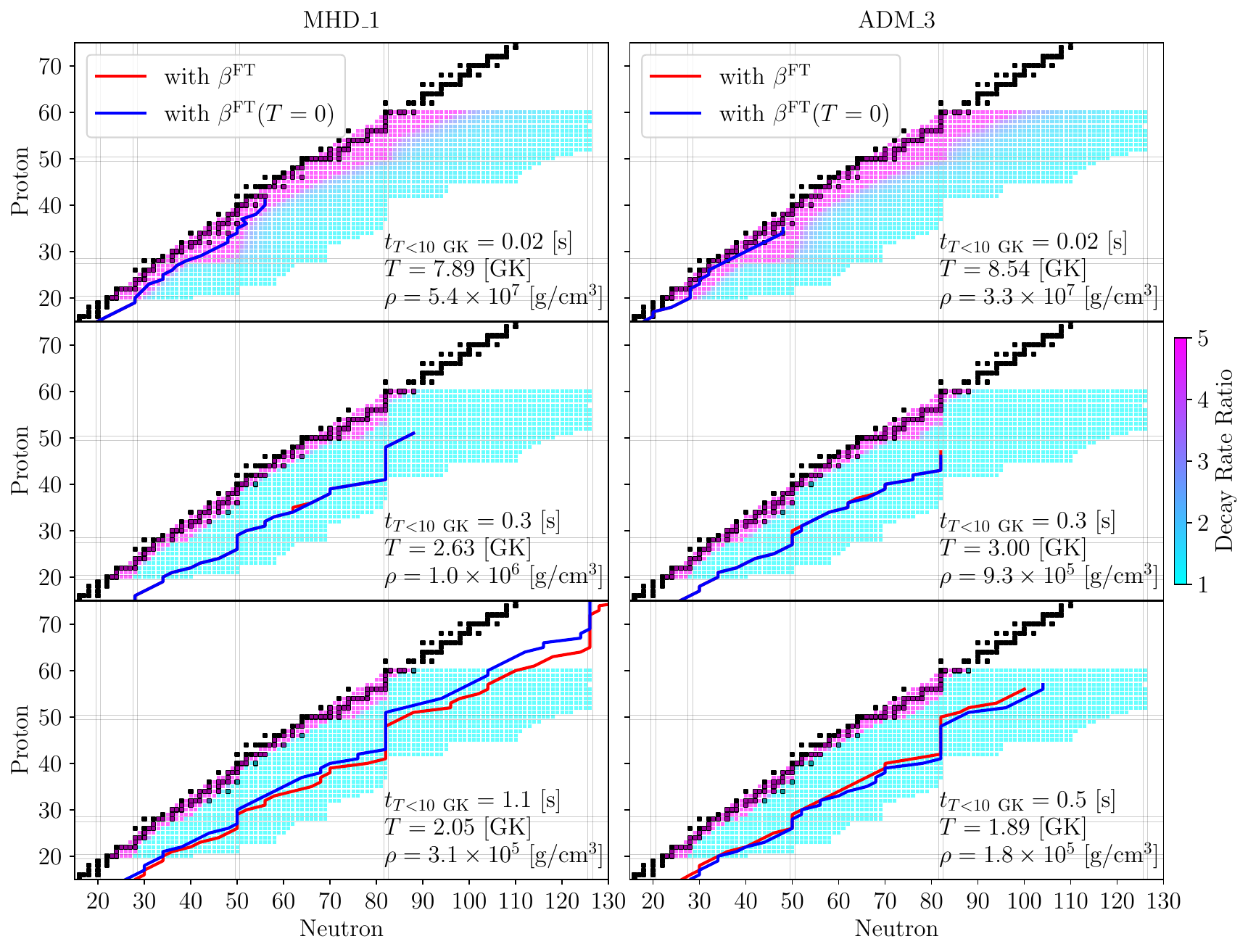}
\caption{
Ratios of the $\beta$-decay rates at finite temperature to those at zero temperature are illustrated on the chart of nuclides at different times for the MHD\_1 (left column) and ADM\_3 trajectories (right column). The solid lines show the path of nucleosynthesis, indicating the nuclei with maximum abundances in each isotopic chain. Red color indicates calculations with temperature-dependent $\beta$-decay rates ($\beta^{\mathrm{FT}}$), while blue color represents zero-temperature rates ($\beta^{\mathrm{FT}}(T=0)$).
\label{fig:b_rate-path}}
\end{figure*}

The effect of $\beta^{\mathrm{FT}}$-decay comes from the accelerated rates at the beginning of the neutron capture process, when the temperature is sufficiently high (close to 10~GK) and the path of nucleosynthesis still lies close to stability. Figure~\ref{fig:b_rate-path} shows the ratio of the $\beta^{\mathrm{FT}}$-decay rates to the zero-temperature rates and the nucleosynthesis paths (solid lines), defined as the abundance maxima in each isotopic chain. The time $t_{T<10~\mathrm{GK}}$ is that elapsed since the temperature has dropped below 10~GK. In the upper panels of the figure, which correspond to the early phase of the nucleosynthesis, it can be seen that the paths in both astrophysical conditions pass through the region of accelerated $\beta$-decay. Subsequently, the temperature drops and the paths move away from stability into the more neutron-rich region, and the finite-temperature effect is no longer significant, as shown in the middle panels of the figure.

The lower panels of Figure~\ref{fig:b_rate-path} show the rate ratios and nucleosynthetic paths at the time when the relative abundances of neutrons drop to $10^{-2}$, roughly corresponding to the onset of neutron freeze-out, when the populated neutron-rich nuclei decay back to stability. A notable difference between the two astrophysical trajectories is that, in MHD\_1, the synthesized material computed with the zero-temperature rates starts to decay back to stability earlier than those computed with the temperature-dependent rates, whereas the opposite happens in the ADM\_3 trajectory.  

\begin{figure*}[ht!]
\includegraphics[width=\textwidth]{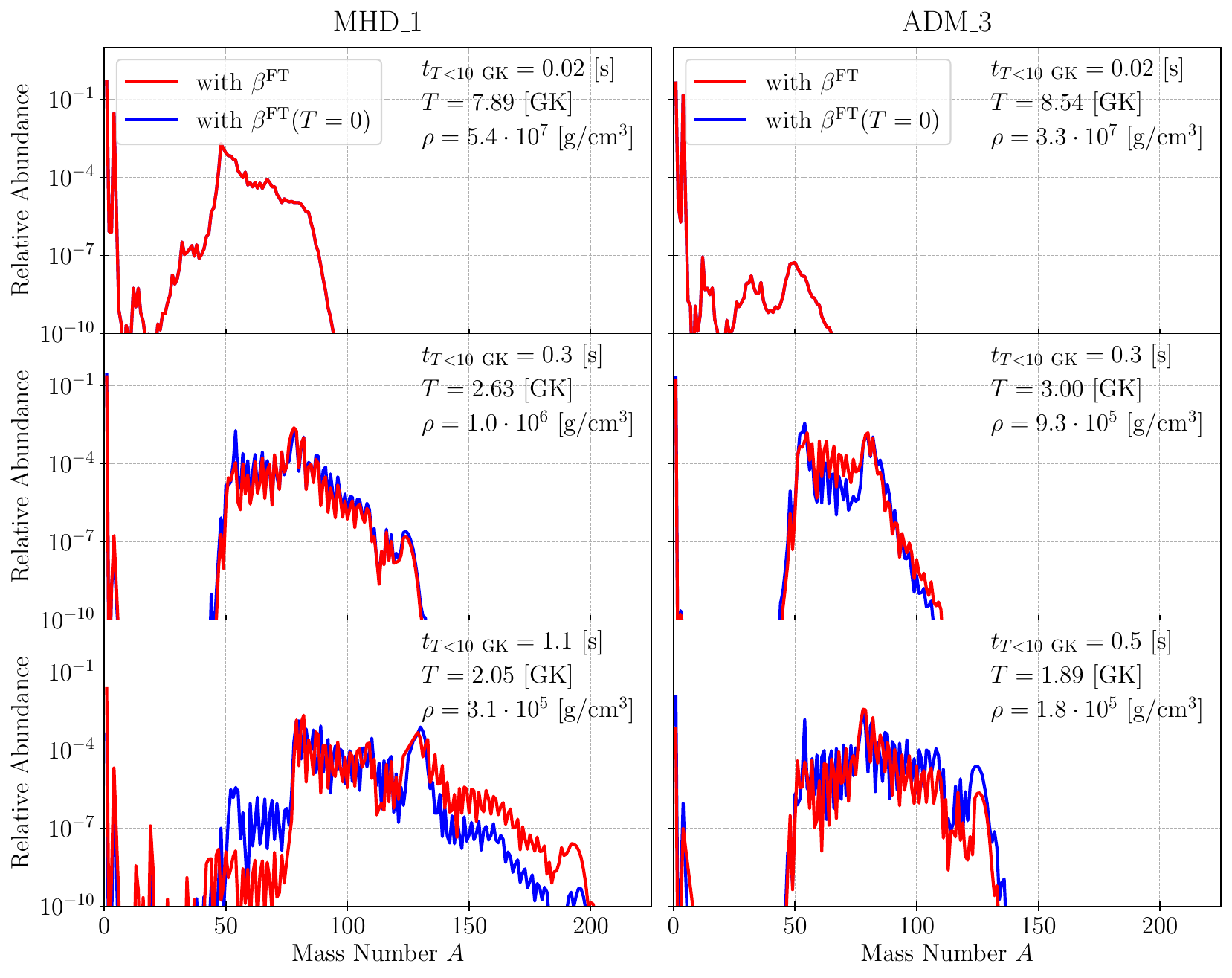}
\caption{
Isotopic abundance patterns (abundances as functions of the mass number $A$) at the times corresponding to Figure~\ref{fig:b_rate-path} for the MHD\_1 (left) and ADM\_3 trajectories (right). Red color indicates calculations with temperature-dependent $\beta$-decay rates ($\beta^{\mathrm{FT}}$), while blue color represents zero-temperature rates ($\beta^{\mathrm{FT}}(T=0)$).
\label{fig:YA_evolution}}
\end{figure*}

\begin{figure*}[ht!]
\includegraphics[width=\textwidth]{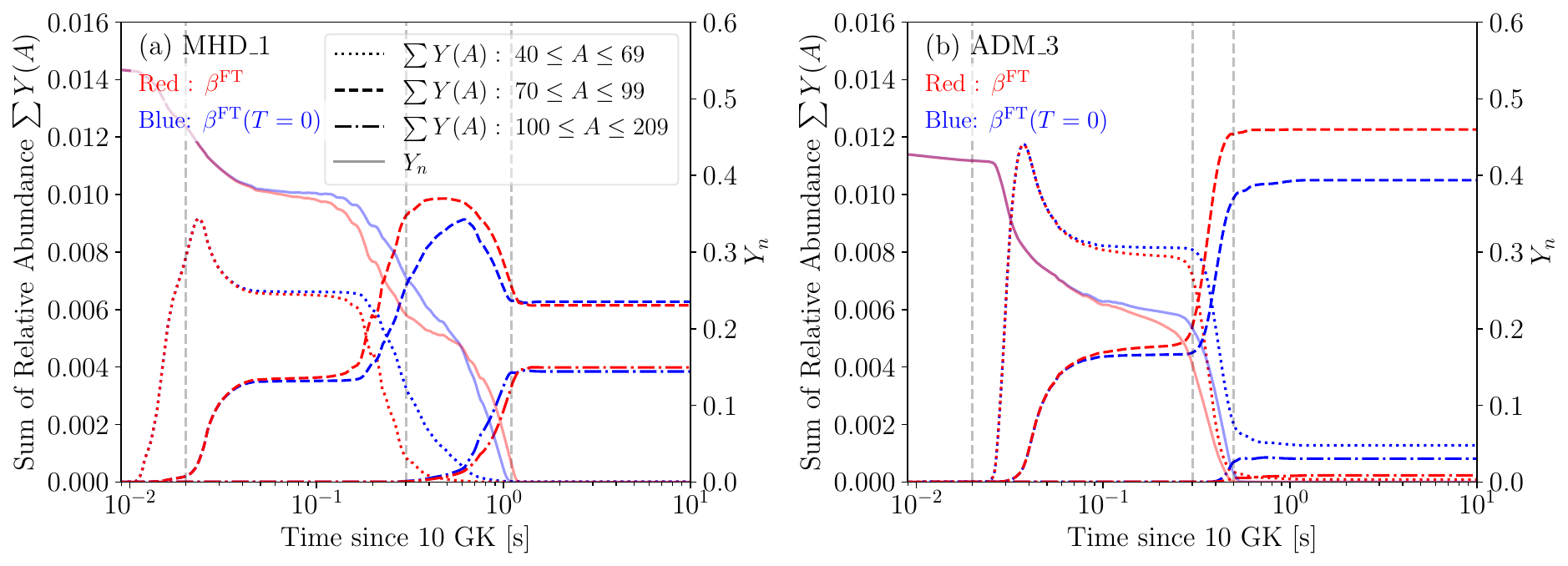}
\caption{
The sums of isotopic abundances are shown as functions of time for nuclei with $40 \leq A \leq 69$ (dotted lines), $70 \leq A \leq 99$ (dashed lines), and $100 \leq A \leq 209$ (dash-dotted lines), for the MHD\_1 (left) and ADM\_3 (right) trajectories. The solid lines show the evolution of the neutron abundance $Y_n$. Red color indicates calculations with temperature-dependent $\beta$-decay rates ($\beta^{\mathrm{FT}}$), while blue color represents zero-temperature rates ($\beta^{\mathrm{FT}}(T=0)$).
\label{fig:Ysum-Yn}}
\end{figure*}

Figure~\ref{fig:YA_evolution} shows the evolution of the isotopic abundance patterns corresponding to each of the panels in Figure~\ref{fig:b_rate-path}. To understand the differences between the calculations with the zero-temperature rates and the temperature-dependent rates, as well as between the two trajectories, the sums of abundances of nuclei with different ranges of mass numbers are shown in Figure~\ref{fig:Ysum-Yn} as functions of time. The solid lines in Figure~\ref{fig:Ysum-Yn} show the neutron abundances as a function of time.

At $t=0.02$~s when the nucleosynthesis paths lie in the region of accelerated rates, the effect of the temperature-dependent rates is not yet visible, as shown in the top panels of Figure~\ref{fig:YA_evolution} and the left most vertical dashed lines in Figure~\ref{fig:Ysum-Yn}. As the nuclei in the $70 \leq A \leq 99$ region are populated and those in the $40 \leq A \leq 69$ region are consumed, the effect of the accelerated $\beta$-decay rates becomes apparent. The faster $\beta$-decay enhances the transfer of abundances from $A\sim 50$ to $A \sim 80$, as shown in the middle panels of Figure~\ref{fig:YA_evolution}, corresponding to the middle vertical dashed lines in Figure~\ref{fig:Ysum-Yn} at $t=0.3$~s. Due to faster synthesis of the $A \sim 80$ nuclei, neutrons are consumed at a faster rate, leading to a decrease in neutron abundances (solid lines in Figure~\ref{fig:Ysum-Yn}) in the case of temperature-dependent rates compared to zero-temperature rates. 

The crucial difference between the two astrophysical conditions arises as follows. In MHD\_1, the abundances of nuclei in the $70 \leq A \leq 99$ region reach a maximum and the heavier nuclei are synthesized using the remaining neutrons. On the other hand, in ADM\_3, neutrons are almost exhausted before nuclei with $A \geq 100$ can be synthesized. This determines whether the temperature-dependent rates enhance the heavier tail of the abundance patterns or not. 

In the case of MHD\_1, while faster synthesis of nuclei in the $70 \leq A \leq 99$ region with the temperature-dependent rates leads to faster neutron consumption at the beginning, the rate of decrease in neutron abundance slows significantly as the abundance of nuclei with $70 \leq A \leq 99$ reaches a maximum, at which point nuclei with $A \geq 100$ start to populate. The combination of larger abundances of $A\sim 80$ nuclei and the extended neutron availability compared to the case with the zero-temperature rates finally leads to the enhancement of the abundances of $A \geq 100$ nuclei. In contrast, for the ADM\_3 trajectory, while the synthesis of the $A \sim 80$ nuclei is similarly enhanced, the neutrons are exhausted before the $A \sim 80$ can reach a maximum and begin populating heavier nuclei. Therefore, in this case, neutrons are mainly consumed in the production of $A \sim 80$ nuclei, leading to the suppression of abundances of heavier nuclei.

The speed with which heavier nuclei with $A \geq 100$ are produced in relation to the evolution of the neutron abundance also depends on the initial composition at 10~GK under NSE. The entropy of the system under an NSE condition is proportional to $T^3 / \rho$ \citep{Cowan2021}. Therefore, comparing the initial densities $\rho$ at 10~GK (Table~\ref{tab:trajectory}) for MHD\_1 and ADM\_3, the initial entropy of MHD\_1 is smaller than that of ADM\_3. Smaller entropy means that more free neutrons and protons are combined into nuclei \citep{Meyer1994}. Thus, the initial abundances of light nuclei are higher in MHD\_1, a sign of which can be seen in the earlier production of $40 \leq A \leq 69$ compared to ADM\_3, as shown in Figure~\ref{fig:Ysum-Yn}. The neutron richness or, equivalently, the electron fraction $Y_e$ also affects the outcome of nucleosynthesis. In addition to the larger initial number of neutrons in MHD\_1, the temperature is sustained longer above $\sim 2$~GK, as shown by $t_{<10~\mathrm{GK}}$ in the lower panels of Figure~\ref{fig:YA_evolution}. This promotes the synthesis of heavier nuclei through the $(n, \gamma) \leftrightarrow (\gamma, n)$ equilibrium, compared to ADM\_3.

In summary, in this section, we have pointed out the seemingly contradictory effect of the temperature-dependent $\beta$-decay rates on producing the heavier tail of the weak $r$-process pattern, taking the MHD\_1 and ADM\_3 trajectories as examples. In both trajectories, the paths of nucleosynthesis lie in the region with accelerated $\beta$-decay rates in the early phase. This enhances the synthesis of nuclei around the first peak of the $r$-process at $A\sim 80$. This also significantly modifies the rate of neutron consumption. In MHD\_1, the larger abundance of nuclei in the $70 \leq A \leq 90$ region combined with the longer-lasting neutron availability leads to the enhancement of the heavier tail ($A \geq 100$) of the abundance pattern. On the other hand, in ADM\_3, the accelerated production of $70 \leq A \leq 90$ nuclei leads to earlier exhaustion of neutrons, leading to the suppression of the heavier tail. This trend applies to varying degrees to other MHD and ADM trajectories, respectively.

\subsection{Effect on the main/strong \textit{r}-process abundances \label{subsec:effect_main-r}}
In this section, we discuss the effect of the finite-temperature $\beta$-decay rates in the case of the main/strong $r$-process conditions.  The main/strong $r$-process refers to the production of elements around the third $r$-process peak ($A\sim195$) and beyond in a given astrophysical event.

For the purpose of discussion, we mainly focus on MHD\_3 and ADM\_8 as examples of the main $r$-process trajectories. This is because among the MHD trajectories, MHD\_3 is the only trajectory in which elements in the third $r$-process peak and beyond can be produced. For the ADM trajectories, there are several main/strong $r$-process trajectories, and we choose the trajectory with the smoothest evolution of temperature and density for a clearer presentation of the result. 

\begin{figure*}[ht!]
\includegraphics[width=\textwidth]{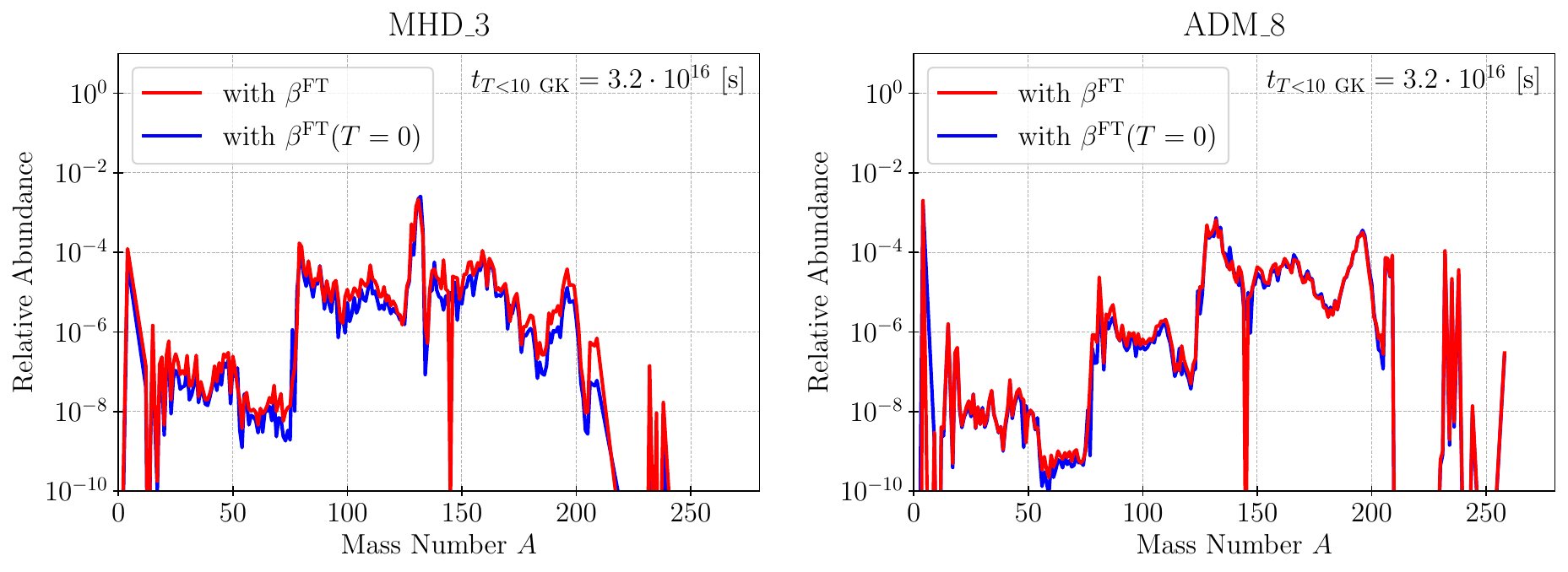}
\caption{
Isotopic abundance patterns (abundances as functions of the mass number $A$) at 1 Gyr to for the MHD\_3 (left) and ADM\_8 (right) trajectories, corresponding to the main/strong $r$-process conditions. Red color indicates calculations with temperature-dependent $\beta$-decay rates ($\beta^{\mathrm{FT}}$), while blue color represents zero-temperature rates ($\beta^{\mathrm{FT}}(T=0)$).
\label{fig:YA_final_main}}
\end{figure*}

Figure~\ref{fig:YA_final_main} shows the final isotopic abundance patterns for the MHD\_3 (left) and ADM\_8 (right) trajectories. We first point out that the dip in the abundance patterns at $A=146$ is due to the non-zero $\beta$-decay half-life of $^{146}$Nd at finite temperature, predicted by the current theoretical $\beta$-decay calculation, although the nucleus is observed to be stable at zero temperature. This occurs because in the abundance calculation the temperature only asymptotically approaches but never reaches zero. 

\begin{figure*}[ht!]
\includegraphics[width=\textwidth]{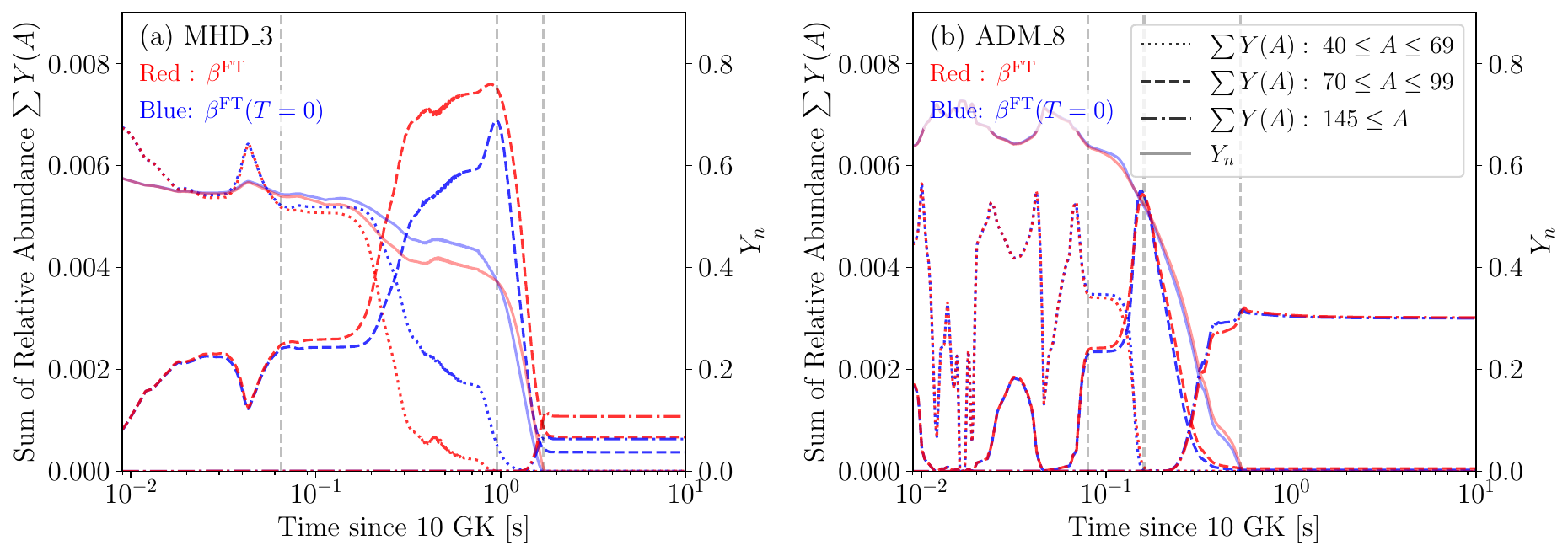}
\caption{
The sums of isotopic abundances are shown as functions of time for nuclei with $40 \leq A \leq 69$ (dotted lines), $70 \leq A \leq 99$ (dashed lines), and $145 \leq A$ (dash-dotted lines), for the MHD\_3 (left) and ADM\_8 (right) trajectories. The solid lines show the evolution of the neutron abundance $Y_n$. Red color indicates calculations with temperature-dependent $\beta$-decay rates ($\beta^{\mathrm{FT}}$), while blue color represents zero-temperature rates ($\beta^{\mathrm{FT}}(T=0)$).
\label{fig:Ysum-Yn_main}}
\end{figure*}

In the MHD\_3 trajectory, it can be seen that the abundance well above the first $r$-process peak at $A\sim80$ is enhanced in the calculation with the finite-temperature $\beta$-decay rates, especially around the third $r$-process peak at $A\sim195$. This effect can be explained in a similar way to the case of the weak $r$-process, as shown in Figure~\ref{fig:Ysum-Yn_main}. The left panel of the figure shows the evolution of the sums of the abundances below the first peak (dotted line), around the first peak (dashed line), and above the second peak (dash-dotted line) in the MHD\_3 trajectory. The figure shows that the accelerated $\beta$-decay rates enhance the consumption of the nuclei below the first peak and increase the abundance around the first peak. As the abundance below the first peak is exhausted, the consumption of neutron slows down, creating the excess of neutrons compared to the calculation with the zero-temperature $\beta$-decay rates. This excess neutron helps synthesize the nuclei in the main/strong $r$-process region.

On the other hand, there is little visible difference in the calculated abundance patterns in the case of the ADM\_8 trajectory. Although the evolution of the abundances shown in the right panel in Figure~\ref{fig:Ysum-Yn_main} indicates a slight effect of the temperature-dependent $\beta$-decay rates, the size of the effect is much smaller than in the case of MHD\_3. This is most likely due to the small amount of time that the path of nucleosynthesis spends in the region with accelerated $\beta$ decay rates at the beginning of the trajectory. 

\begin{figure}[ht!]
\includegraphics[width=0.5\textwidth]{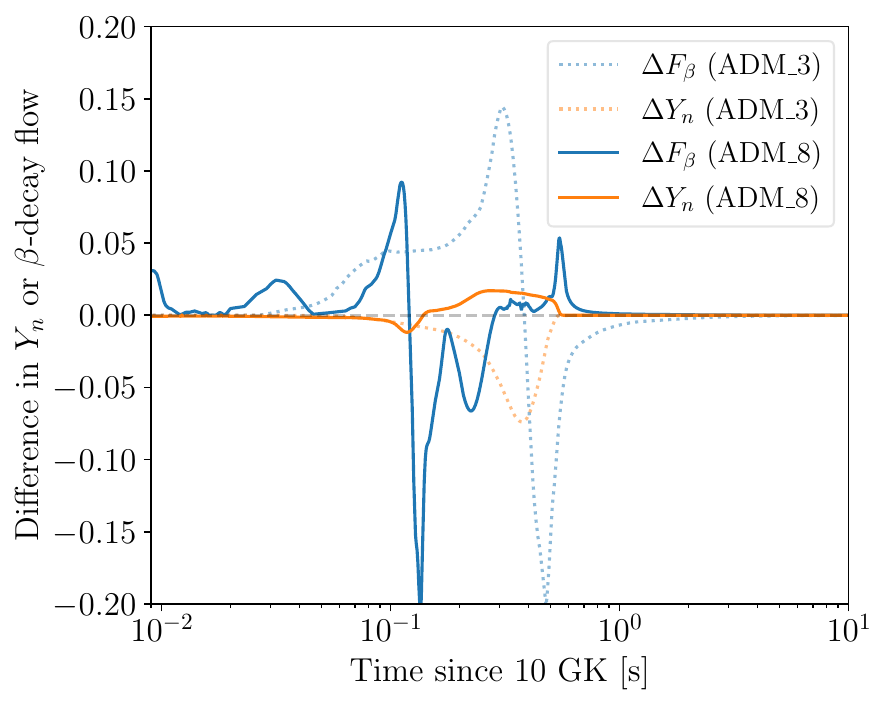}
\caption{
The differences in the total flows of $\beta$-decays ($\Delta F_{\beta}$) and neutron abundances ($\Delta Y_n$) as functions of time, between the calculations with the temperature-dependent and zero-temperature $\beta$-decay rates\label{fig:Yn-beta-diff}.}
\end{figure}

This is illustrated in Figure~\ref{fig:Yn-beta-diff}, which shows the difference in the total flows of $\beta$-decays ($F_{\beta}$) and neutron abundances ($Y_n$) as functions of time, between the calculations with the temperature-dependent and zero-temperature $\beta$-decay rates:
\begin{align}
    \Delta F_{\beta}(t) &= F_{\beta}^{\mathrm{FT}}(t) - F_{\beta}^{\mathrm{FT} (T=0)}(t), \\
    \Delta Y_n(t) &= Y_n^{\mathrm{FT}}(t)  - Y_n^{\mathrm{FT}(T=0)}(t),
\end{align}
where ``FT'' means that the calculation was performed with the temperature-(electron density)-dependent $\beta$-decay rates, and ``FT$(T=0)$'' means that the calculation was performed with the $\beta$-decay rates fixed to the values at $T=0$ and $\rho Y_e =1$.
The total flow of $\beta$-decays at time $t$, $F_{\beta}(t)$,  is calculated as 
\begin{align}
    F_{\beta}^{\mathrm{FT}}(t) = \sum_{(N,Z)}&\lambda_\beta^{(N,Z)} (T(t), \rho Y_e(t)) Y^{(N,Z)}(t), \\
    F_{\beta}^{\mathrm{FT}(T=0)}(t) = \sum_{(N,Z)}&\lambda_\beta^{(N,Z)} (T=0, \rho Y_e=1) Y^{(N,Z)}(t), 
\end{align}
where $\lambda_\beta^{(N,Z)}$(t) and $Y^{(N,Z)}(t)$ are the $\beta$-decay rate and the abundance of the species $(N,Z)$ at time $t$, respectively. In this case, the sum is for all the species in the reaction network.
In Figure~\ref{fig:Yn-beta-diff}, the initial increase in $\Delta F_{\beta}$ above zero indicates the acceleration of $\beta$-decay. Compared to the weak $r$-process trajectory (ADM\_3, dotted lines), the enhancement of the $\beta$-decay flow in the main/strong trajectories (ADM\_8, solid lines) in the early phase of the nucleosynthesis is smaller and shorter. Note that the horizontal axis of Figure~\ref{fig:Yn-beta-diff} is shown in log scale. This results in a much smaller effect on the neutron abundance, therefore, the overall effect on the abundance pattern is also limited. A similar trend is observed in other ADM main/strong trajectories.

Another notable effect seen in Figure~\ref{fig:Yn-beta-diff} is that the initial increase in relative $\beta$-decay flows ($\Delta F_{\beta}$) is followed by a dip, in both ADM\_3 and ADM\_8. These dips occur shortly after the relative decrease in the relative neutron abundances $\Delta Y_n$. This is because, initially, consumption of neutrons is enhanced by the faster temperature-dependent $\beta$-decay rates. The relative excess in the neutron abundance in the calculations with the zero-temperature $\beta$-decay rates, signified by the dip of $\Delta Y_n$ below zero, momentarily pushes the path of nucleosynthesis toward a more neutron-rich region, compared to the calculations with the finite-temperature $\beta$-decay rates. The average $\beta$-decay rates are faster in the neutron-rich region; therefore, $\Delta F_\beta$ also dips below zero. As illustrated in this example, not only is the flow of $\beta$-decay directly affected by the temperature dependence of the $\beta$-decay rates, the flow is also subsequently affected by the modification of the path of nucleosynthesis.

\subsection{Effect on the overall \textit{r}-process abundances \label{subsec:effect_overall-r}}
So far we have looked at the effect of the temperature-dependent $\beta$-decay rates on individual trajectories. In this section, 
we explore the sum of the final abundance patterns in both astrophysical conditions, respectively. In the case of the MHD trajectories, the ten trajectories were selected by the authors of \citep{Reichert2021} so that the sum approximates the integrated final abundance pattern of all tracers (trajectories) from the original hydrodynamical simulations. For ADM, these trajectories were selected using the ADM method \citep{Holmbeck+2023} so that the sum of the final abundance patterns approximately reproduces the observed $r$-process abundance pattern in metal-poor stars.
Therefore, the sum of the final abundance patterns of all trajectories is a meaningful quantity for both astrophysical conditions.

\begin{figure*}[ht!]
\includegraphics[width=\textwidth]{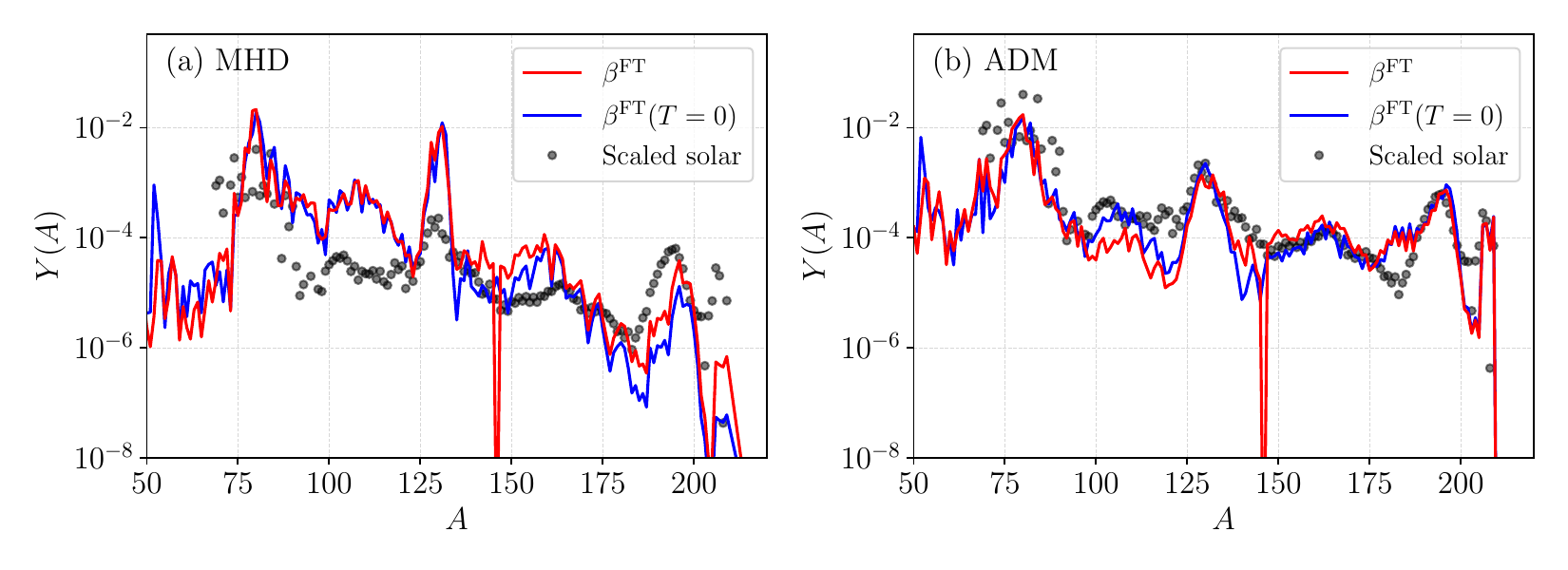}
\caption{
The sums of final abundance patterns of all trajectories for MHD (left) and ADM (right), comparing the calculations with zero temperature (red) and temperature dependent (blue) $\beta$-decay rates. Solar $r$-process abundance pattern (black circles), scaled to match the abundance calculated with $\beta^\mathrm{FT}$ at $A=165$, is also shown for reference.
\label{fig:MHD_ADM_sum_YA}}
\end{figure*}

Figure~\ref{fig:MHD_ADM_sum_YA} shows the sums of the final abundance patterns for all trajectories in MHD (left) and ADM (right), comparing the calculations with the zero temperature (blue) and temperature-dependent $\beta$-decay rates (red). The derived solar abundance pattern \citep{Goriely1999_solar} is also shown for reference, scaled to match the abundance of the abundance pattern calculated with the temperature-dependent rates at $A=165$. For the sum of abundances in MHD, the effect of the temperature-dependent rates is visible above the second $r$-process peak at $A\sim130$. In particular, the relative abundances are also enhanced around $A\sim150$ and $A>175$. For ADM, the effect is most noticeable for $100<A<175$. The difference between the calculations with the two sets of $\beta$-decay rates in the ADM condition suggests that, overall, the temperature-dependent rates enhance the consumption of the abundances below the second $r$-process peak, and increase the production of elements near the rare-earth peak around $A\sim165$. As discussed for some of the specific trajectories in the previous sections, the effect depends on various conditions such as the neutron-richness of the environment and how long the nucleosynthesis path lies in the region of accelerated $\beta$-decay rates. Overall, the effect of temperature-dependent $\beta$-decay rates past the second $r$-process is more visible in the MHD condition. This is because the main/strong $r$-process trajectories in the ADM condition are significantly more neutron-rich ($Y_e < 0.15$) as listed in Table~\ref{tab:trajectory}, therefore the path of nucleosynthesis spends little time near stability where the finite-temperature effect on $\beta$-decay rates is larger.

\subsection{Effect on the energy release}
In $r$-process nucleosysnthesis, heating (energy release) from various nuclear decays and reactions is also of interest, in addition to the resulting abundance patterns. The released energy powers the electromagnetic emission from the astrophysical event on a time scale of days to weeks, which may provide unique evidence of the $r$-process. On a shorter time scale, in the early phase of the $r$-process when the baryon density is still high, the energy release heats the material and may significantly alter the temperature evolution, consequently affecting the nucleosynthesis yields. This is known as nuclear (re)heating.

\begin{figure*}[ht!]
\includegraphics[width=\textwidth]{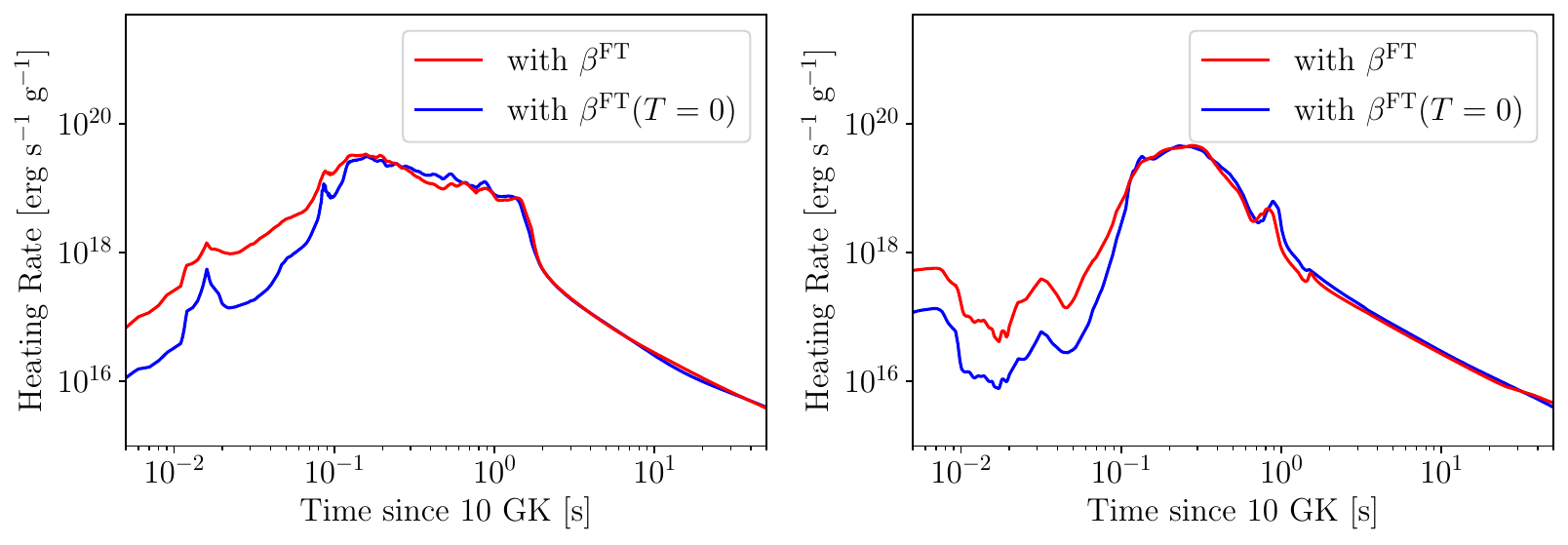}
\caption{
Heating rate (energy release) from all nuclear decays and reactions, summed over all trajectories for MHD (left), and ADM (right), comparing the calculations with zero temperature (red) and temperature dependent (blue) $\beta$-decay rates. The horizontal axis is defined as the elapsed time since the temperature drops below 10~GK. Therefore, it does not reflect the timestamp of each tracer particle. See text for details.
\label{fig:heating_rate_MHD_ADM}}
\end{figure*}

Figure~\ref{fig:heating_rate_MHD_ADM} shows the effect of temperature-dependent $\beta$-decay rates on the summed heating (energy release) rate for the MHD (left) and ADM (right) trajectories. Here, the time origin ($t=0$ [s]) of each trajectory is aligned with the point when the temperature drops below 10 GK. At this temperature, the system is still in nuclear statistical equilibrium (NSE) and it is when our reaction network calculations are started. Therefore, it does not reflect the temporal evolution of tracer particles of the original astrophysical simulations. Rather, the purpose of this figure is to illustrate how the heating rate is affected by the temperature dependence of the $\beta$-decay rates. As expected from previous discussions, the acceleration of the $\beta$ decay rates from the finite-temperature effect significantly increases the amount of energy released in the early phase of nucleosynthesis.

Significant effects are seen in both astrophysical conditions before the heating rates reach their peaks. Considering that the time scale of interest for kilonova is days, this effect most likely does not have direct observable consequences. 

However, an increase in the heating rate in the early stage of nucleosynthesis may have an effect on the temperature evolution of the environment and alter the final abundance pattern. 
{To investigate the effect of the energy release on the resulting abundance patterns, we perform nuclear reaction network calculations including nuclear heating with heating efficiency $\epsilon =$ 0.5 and 1.0. The treatment of nuclear heating closely follows that of Appendix A of \cite{Lippuner2017}.}

\begin{figure*}[ht!]
\includegraphics[width=\textwidth]{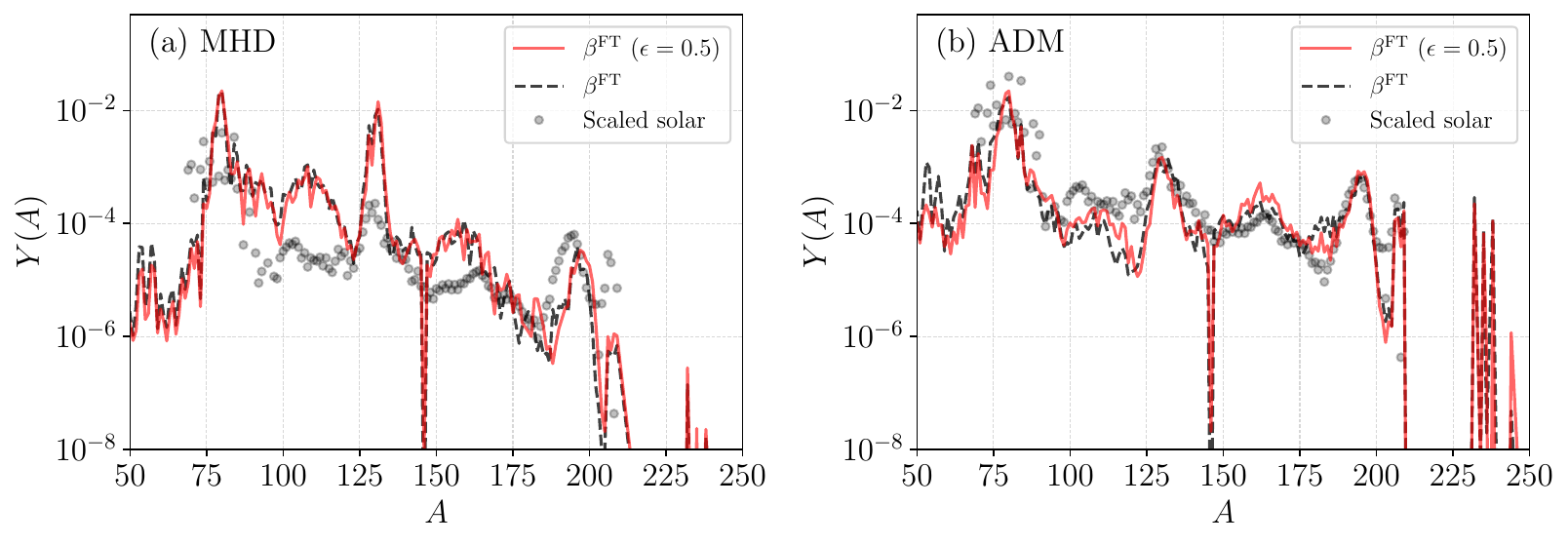}
\caption{
Comparison of the final summed abundance patterns with (red solid lines) and without nuclear heating (black dashed lines), for the MHD trajectories (panel a) and the ADM trajectories (panel b).
\label{fig:YA_heating}}
\end{figure*}

{Figure~\ref{fig:YA_heating} shows the comparison of the final summed abundance patterns with (red solid lines) and without nuclear heating (black dashed lines), for the MHD trajectories (panel a) and the ADM trajectories (panel b). Only the results with heating efficiency $\epsilon = 0.5$ are shown since the differences between $\epsilon = 0.5$ and 1 are small. For both trajectories, it can be seen that the inclusion of nuclear heating lowers the abundances below $A\sim75$, due to the increase in temperature at an early time, which further accelerates $\beta$-decay. For the MHD trajectories, the third $r$-process peak is slightly shifted towards the larger mass number, and the abundances above the peak are also slightly enhanced. In the case of the ADM trajectories, the most significant effect is the increase in the height of the rare earth peak. A slight increase in the height of the third $r$-process peak and in the abundance just below $A\sim250$ is also present. In general, the inclusion of nuclear heating moderately enhances the acceleration of $\beta$-decay and further moves the abundances from below $A\sim75$ to higher mass numbers, the extent of which depends on the astrophysical condition. }

{
We note that, in most hydrodynamical simulations of astrophysical events, an approximate effect of nuclear reheating is already included by estimating nuclear abundances assuming nuclear statistical equilibrium (NSE) or by a small reaction network. When a significant contribution from shock or viscous heating is present, careful subtraction of the approximated nuclear heating would be required to avoid double counting when adding nuclear heating using a full reaction network. The results shown here assume that shock or viscous heating is not negligible.
}

\section{Conclusion\label{sec:conclusion}}
In this paper, we explored the effect of theoretical temperature-(electron density)-dependent $\beta$-decay rates of nuclei with proton numbers $20 \leq Z \leq 60$ on the $r$-process nucleosynthesis, through the reaction network calculations. The $\beta$-decay rates were calculated using the FT-PNRQRPA framework. For astrophysical conditions, we considered a set of thermodynamical trajectories from the magnetohydrodynamic (MHD) simulation of a magnetorotational supernova, and the outflow from the disk surrounding a hypermassive neutron star (HMNS), where the set of trajectories was collected with the Actinide Dilution with Matching (ADM) method.

The importance of $\beta$-decay in the $r$-process has typically been discussed in the literature, focusing on the $(n, \gamma) \leftrightarrow (\gamma, n)$ phase and the freeze-out phase. However, its impact in the early stage of the $r$-process has rarely been discussed. In this study, it was shown that in the early phase of the $r$-process when the temperature is large ($T>5$~GK), the path of nucleosynthesis (abundance maxima for each isotopic chain) is still close to stability and passes through the region where the $\beta$-decay rates are faster due to the finite temperature effect. This affects the subsequent abundance evolution, depending on the neutron-richness and temperature evolution of the astrophysical conditions.

In weak $r$-process conditions, the accelerated $\beta$-decay rates enhance the synthesis of the elements in the first $r$-process peak region ($A\sim 80$). After this point, depending on the availability of free neutrons, further synthesis of heavier elements is either enhanced (e.g. MHD\_1) or suppressed (e.g. ADM\_3).

In the main/strong $r$-process condition within the set of MHD trajectories, namely MHD\_3, the accelerated $\beta$-decay rates have a similar effect of enhancing the production of the first $r$-process peak nuclei, and eventually enhance the production of the third $r$-process peak nuclei and beyond. On the other hand, when the neutrons are consumed much more quickly as in ADM\_8, the impact of the accelerated $\beta$-decay rates on the final abundance pattern is limited.

The inclusion of temperature-dependent $\beta$-decay rates also impacts calculations of nuclear heating. Due to the faster $\beta$-decay rates, when the temperature is high and the path of nucleosynthesis is close to stability, we note increased energy release in both sets of astrophysical conditions considered. We demonstrate that this increased nuclear heating results in additional changes to the final abundance patterns, though these changes are modest. However, we emphasize that the temperature-dependent $\beta$-decay rates explored here are available only for a limited region of the nuclear chart. We look forward to future work in which we will extend our studies to include the relevant temperature-dependent $\beta$-decay rates of both weak and main $r$-process nuclear species. 

\section{Acknowledgments}

Y.S. and R.S. acknowledge support from the U.S. National Science Foundation under grant number 21-16686 (NP3M). This work is supported in part by the U.S. Department of Energy under contracts DE-SC00268442 (SciDAC ENAF---R.S. and P.N.), DE-FG02-95-ER40934 (R.S.), and DOE-DE-NA0004074 (NNSA, the Stewardship Science Academic Alliances program---A.R.). P.N. acknowledges support from the Arthur J. Schmitt Foundation. This research was supported in part by the Notre Dame Center for Research Computing through its computing infrastructure.





\bibliography{main}{}
\bibliographystyle{aasjournal}



\end{document}